\documentclass[aps,prb,floatfix,twocolumn,,a4paper]{revtex4}
\usepackage{amsmath}
\usepackage{amsfonts}
\usepackage{graphicx}
\usepackage{epsfig}
\usepackage{amssymb}
\usepackage{color}

\begin{document}

\title{Quantum Down Conversion and Multipartite Entanglement via a
Mesoscopic SQUID Ring}
\author{P.B. Stiffell}
\author{M.J. Everitt}
\author{T.D. Clark}
\email{t.d.clark@sussex.ac.uk}
\author{C.J. Harland}
\affiliation{Centre for Physical Electronics and Quantum Technology, School of
Engineering, University of Sussex, Brighton, Sussex BN1 9QT, U.K.}
\author{J.F. Ralph}
\affiliation{Liverpool University, Brownlow Hill, Liverpool, L69 3GJ, U.K.}
\date{22 nov 2004}

\begin{abstract}
In this paper we study, by analogy with quantum optics, the SQUID ring
mediated quantum mechanical interaction of an input electromagnetic field
oscillator mode with two or more output oscillator modes at sub-integers of
the input frequency. We show that through the non-linearity of the SQUID
ring multi-photon down conversion can take place between the input and
output modes with the resultant output photons being created in an entangled
state. We also demonstrate that the degree of this entanglement can be
adjusted by means of a static magnetic flux which controls the strength of
the interaction between these modes via the SQUID ring.
\end{abstract}

\maketitle

\section{Introduction}

Over the last five years there has been a rapidly growing interest in
quantum technologies, particularly for the fields of quantum information
processing and quantum computing~\cite{lo,bouwmeester}. This has been aided
both by the creation of key algorithms~\cite{shor,grovner} and by
improvements in experimental technique. A range of approaches, involving
different physical disciplines, have been adopted in the pursuit of these
technologies but at the basis of each is the concept of using entangled
states of quantum objects~\cite{lo,bouwmeester}. The creation of entangled
states, and the exploitation of their properties, are therefore seen as
fundamental to successful progress in these technologies. In particular, it
is clear that there exists a need to develop systems or devices through
which the degree of entanglement between quantum objects can be actively
controlled. In this paper we show how the intrinsic non-perturbative
behaviour of a quantum SQUID ring (here, a thick superconducting ring
enclosing a Josephson weak link device) can be used to achieve this goal,
taking as our specific example entanglement between photon states.

Interest in the use of SQUID rings (and other weak link circuits) as
quantum logic devices has been stimulated by recent experiments to
probe superposition of states in these
systems~\cite{wal,friedman,nakamura,nakamura99,lupascu,tim}. Furthermore,
other recent experiments have been performed that show coherent
dynamics of superconducting flux qubits coupled to individual harmonic
oscillators~\cite{Chiorescu2004}. Building on this background, and
previous theoretical work by us on the coupling of SQUID rings to
electromagnetic (em) fields~\cite{mje}, we consider the process of
photon down conversion via a mesoscopic SQUID ring. Specifically, by
analogy with the field of quantum optics, we demonstrate that the ring
may be viewed as a non-linear medium which can down convert photons
from one frequency (input field oscillator mode) to another lower
frequency and generate entangled photons (output field oscillator
modes), as is often performed in quantum optics
experiments~\cite{scully}. An analysis of multiple output modes is
important since it allows us to understand the way in which several
quantum circuits may be entangled - an important issue in quantum
technologies - for example, as entanglement registers in quantum
computing. We also note that by coupling in a larger number of field
modes we change the position (in $\Phi_{x}$) of the splittings in the
spectrum of the Hamiltonian for the composite system. We consider,
therefore, that it is by no means trivial to assume that we can simply
extend the results of our previous work~\cite{mje2,mje}.  As we have
shown by computation, in a three mode system (two em field modes plus
one SQUID mode) entanglement can be adjusted through the control of a
static bias flux $\Phi _{x}$ applied to the ring. From an experimental
viewpoint, in this system down conversion processes are relatively
easy to observe~\cite{jmodopt} at least for the situation where
microwaves (input electromagnetic field $\approx $ in the range of a
few to 10GHz) are down converted to radio frequencies (rf output mode
$\approx $ 1 to 20MHz)~\cite{jmodopt}. We note that microwave photons
are difficult to transport, which could be problematic for some
applications in quantum communication.  By implication, but not
calculated in this work, it should also prove possible to observe up
conversion processes involving significant frequency ratios between
input and output. However, the ultra low noise electronics required
for the output stage in an experiment of this type (say, for inputs at
rf and outputs at microwave frequencies) is much more difficult to
engineer than at the much lower radio frequencies. Having observed
down conversion in the experiments at Sussex~\cite{jmodopt}, we tried
to emulate this behaviour theoretically. However, such large frequency
differences (a few GHz input, a THz SQUID ring characteristic
oscillator frequency and a few MHz output) was found to be
computationally intractable at this time. Bearing this in mind, and
considering the interest currently shown in quantum computing, we
chose to demonstrate in this paper that SQUID rings can act as devices
which mediate large ratio frequency down conversion that is also
computationally tractable. Moreover, for multiple output modes we
demonstrate that the down converted photons are entangled together.
In our computations we have modelled the field oscillator modes as
equivalent resonant LC circuits. Experimentally, this could be
realised by utilising real LC circuits constructed from
superconducting material to minimise decoherence. Alternatively we
could use a transmission line comprising a series of coupled
superconducting LC resonators~\cite{jmodopt,Wallraff2004}.

From an experimental perspective, it is interesting to note that
there has already been some progress made in realising simple quantum
circuits systems relevant to the work presented in this paper. In this
regard there are now several examples of coupled superconducting
qubits (here the SQUID ring is, to all intents and purposes, a flux
qubit). These include fixed coupling flux qubits~\cite{Majer2005},
current biased Josephson-junction qubits~that are either coupled
capacitively~\cite{Berkley2003} or via a connecting superconducting
loop~\cite{Kim2004}.  Recently, there has even been an example of
tri-partite quantum entanglement in macroscopic superconducting
circuits~\cite{Xu2005}. Moreover, an application that relates very
strongly to the circuit systems presented in this work utilised a
SQUID-ring to mimic a simple harmonic oscillator. This device was then
coupled, quantum coherently, to a superconducting flux
qubit~\cite{Chiorescu2004}. However, although the physics of all these
systems is similar, in this work we have chosen to consider a
superconducting circuit coupled to quantum electrodynamic degrees of
freedom, i.e. photons in a cavity. Recently it has been shown that a
superconducting qubit can be quantum coherently coupled to an on-chip
cavity~\cite{Wallraff2004}.  Furthermore, we note that superconducting
devices, which can be coupled to these and more exotic systems, such
as GHz nanomechanical resonators~\cite{Geller2005}, are not only easy
to fabricate, but are adaptable to a wide variety of applications.

In this work we first investigate, from a full quantum mechanical
viewpoint, the simplest down conversion and entanglement process,
involving one input photon at a given frequency and two output photons
at half this input frequency. To the limits of the computational power
available to us, we extend our theoretical investigation of such down
conversion processes to that of generating four output photons, each
at a quarter of the frequency of the input photon (i.e. a four photon
down conversion). Within the Adami-Cerf
criterion~\cite{cerf,hey,nielsen} we show that entanglement exists
between any one of these output oscillator modes and the rest of the
system (i.e. SQUID ring and input oscillator mode). This implies that
all the output modes will be entangled with each other, as well as to
the rest of the coupled system. While this multipartite entanglement
may not be as strong as some bipartite entanglements, such as Bell
States, it is a key component in certain requirements in quantum
computation, such as entanglement registers, where a large number of
qubits are entangled together to generate a total entangled state for
the process of computation.  It is this idea that is responsible for
the (as yet in principle) power of quantum computing over modern day,
classical computers. Also, the question of characterising entanglement
in multipartite quantum systems is still very much an open one. Whilst
each measure agrees as to whether the state vector of a given system
is separable, or maximally entangled, they do not, in general, provide
an ordering that can be used to quantify entanglement of more general
states. In this paper we have chosen to use an entanglement measure
that gives us an idea of the distribution of the quantum correlations
that exist in our example multicomponent systems. However, as we have
not produced maximally entangled states it is not possible to quantify
absolutely the level of entanglement present. We note that for a
simpler system, controlled via a dynamic flux bias~\cite{mjeArchive},
we have shown that it is possible to create a maximally entangled
state between a SQUID ring and an em-field mode. It may be that
similar methods could be used in these composite systems to produce
such states of maximal entanglement. Unfortunately computations with
time dependent Hamiltonians are far harder to solve than the static
cases presented in this paper, and are beyond the limit of current
computational power.

For the example of the two photon process we consider the effect of
dissipation on the coherent evolution of the coupled (entangled) system. By
extension, we show by computation that much higher order (i.e. more than our
computed four photon case) down conversions may prove possible
experimentally. To this effect we provide a further example of down
conversion by a factor of twenty between the input em oscillator mode and an
output mode. However, both due to limitations in computer power, and a still
insufficient understanding of entanglement in many particle (multipartite)
quantum systems, we have not proceeded any further in characterising this
high order example.

\section{Theoretical model}

In our theoretical treatment of the interaction of quantised em oscillator
modes (photons) with a mesoscopic quantum mechanical SQUID ring we model the
em field modes as parallel LC resonators, as a simple and convenient way of
parameterizing the (physical) cavity modes coupled to the SQUID ring. The
scheme we have adopted is shown diagrammatically in 
\begin{figure}[t]
\begin{center}
\resizebox*{0.45\textwidth}{!}{\includegraphics{{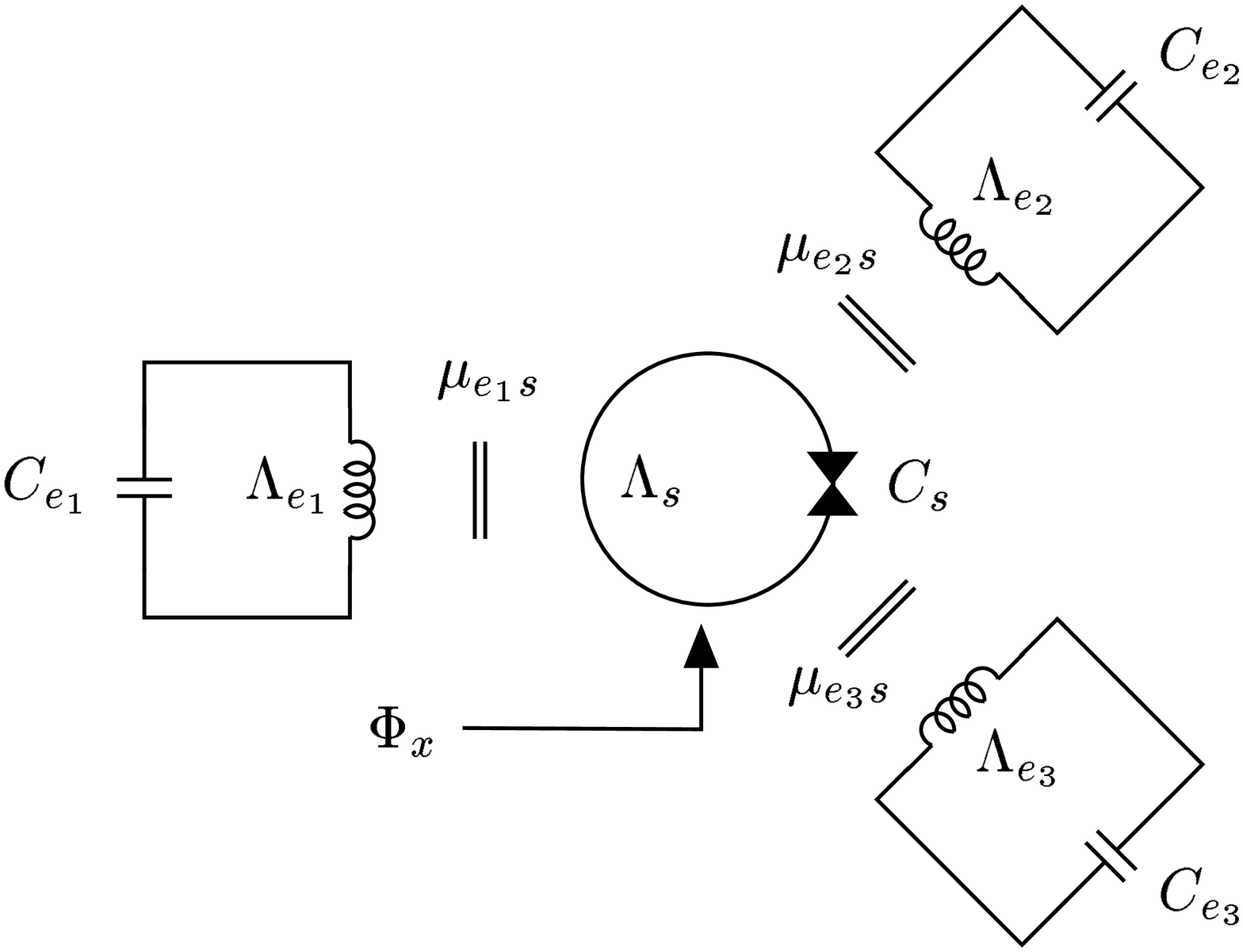}}}
\end{center}
\caption{Schematic of a four mode system comprising three electromagnetic
field mode inductively coupled to a mesoscopic SQUID ring.}
\label{QDCFactor2_Schematic}
\end{figure}
figure~\ref{QDCFactor2_Schematic} for the simpler, two photon, down
conversion process. In this example the input photon frequency is
$\omega _{e_{1}}/2\pi $ while the two (identical) output frequencies
are $\omega _{e_{2}}/2\pi $ and $\omega _{e_{3}}/2\pi $ such that
$\omega _{e_{2}}=\omega _{e_{3}}=0.5\omega _{e_{1}}$. Here, the SQUID
ring characteristic oscillator frequency is $\omega _{s}/2\pi =1/2\pi
\sqrt{\Lambda _{s}C_{s}}$ for a ring inductance and a weak link
capacitance of $\Lambda _{s}$ and $C_{s}$, respectively. In our
calculations we treat this system fully quantum mechanically by
assuming an operating temperature $T$ such that $\hbar \omega
_{s}\left( =\hbar /\sqrt{\Lambda _{s}C_{s}}\right) ,\hbar \omega
_{e_{1}}\left( =\hbar /\sqrt{C_{e_{1}}L_{e_{1}}}\right) $and $\hbar
\omega _{e_{2,3}}\left( =\hbar /\sqrt{C_{e_{2,3}}L_{e_{2,3}}}\right) $
are all much greater than $k_{B}T$ for field mode inductances and
capacitances $L_{e_{i}}$ and $C_{e_{i}}$ $($where, generally,
$i=1,2,3.....)$. In this model system the coupling between the SQUID
ring and the oscillator modes is taken to be inductive. As can be seen
in figure~\ref{QDCFactor2_Schematic}, we have chosen to couple these
circuits so that all interactions between the field modes are
performed via the SQUID ring. In this four mode system (i.e. three em
modes plus one SQUID mode) the static flux $\Phi _{x}$ applied to the
ring can be used to control the level of interaction between the input
em mode and an output em mode of the system.  In a specific example in
this paper, as an extension of previous work~\cite{mje}, we
demonstrate that the behaviour of coupled SQUID ring-em mode systems
is, in general, very strongly dependent on this control flux. Although
this aspect of our investigations is an extension of the results of
our previous papers~\cite{mje2,mje}, it is important to show that this
dependence remains valid for the kind of multicomponent circuits that
would appear to be required in the quantum technologies currently
being
pursued~\cite{lo,bouwmeester,wal,friedman,nakamura,nakamura99,lupascu,tim,Chiorescu2004,nielsen}.

The Hamiltonian for the coupled system of figure~\ref{QDCFactor2_Schematic}
is made up of the uncoupled Hamiltonians for each mode together with a set
of interaction terms. For each em field mode (both input and output) the
Hamiltonian takes the form~\cite{mje} 
\begin{equation}
H_{e_{i}}=\frac{Q_{e_{i}}^{2}}{2C_{e_{i}}}+\frac{\Phi_{e_{i}}^{2}} {2\Lambda_{e_{i}}}  \label{eq:emfield}
\end{equation}
where, again, $i=1,2,3.....$, while for the SQUID ring we use the standard,
time independent, lumped component circuit Hamiltonian~\cite{timplt,allenpr}
(the wavelengths of the photon modes being very much larger than the
dimensions of the mesoscopic SQUID ring circuit). Specifically
\begin{equation}
H_{s}=\frac{Q_{s}^{2}}{2C_{s}}+\frac{\left( \Phi_{s}-\Phi_{x}\right)
^{2} }{2\Lambda_{s}}-\hbar\nu\cos\left(2\pi\frac{\Phi_{s}}{\Phi_{0}}\right)
\label{eq:HamS}
\end{equation}
In this description $\Phi_{s}$ (the magnetic flux threading the SQUID
ring) and $Q_{s}$ (the electric displacement flux between the
electrodes of the weak link) are the conjugate variables for the ring,
$\hbar\nu/2$ is the matrix element for Josephson pair tunnelling
through the weak link (of critical current $I_{c}=2e\nu$) and
$\Phi_{0}=h/2e$ is the superconducting flux quantum. From the
perspective of the work presented in this paper, the Josephson cosine
term in the SQUID ring Hamiltonian~(\ref{eq:HamS}) allows for
non-perturbative behaviour to all orders in the coupled system of
figure~\ref{QDCFactor2_Schematic}. In turn, this means that in
interactions between a quantum SQUID ring and external em fields,
non-perturbative, multiphoton absorption/emission processes tend to
dominate within an exchange region~\cite{mje}.

To illustrate the form of the solutions to the time independent
Schr\"{o}dinger equation for a SQUID ring, using the
Hamiltonian~(\ref{eq:HamS}), we show in
\begin{figure*}[t]
\begin{center}
\resizebox*{0.8\textwidth}{!}{\includegraphics{{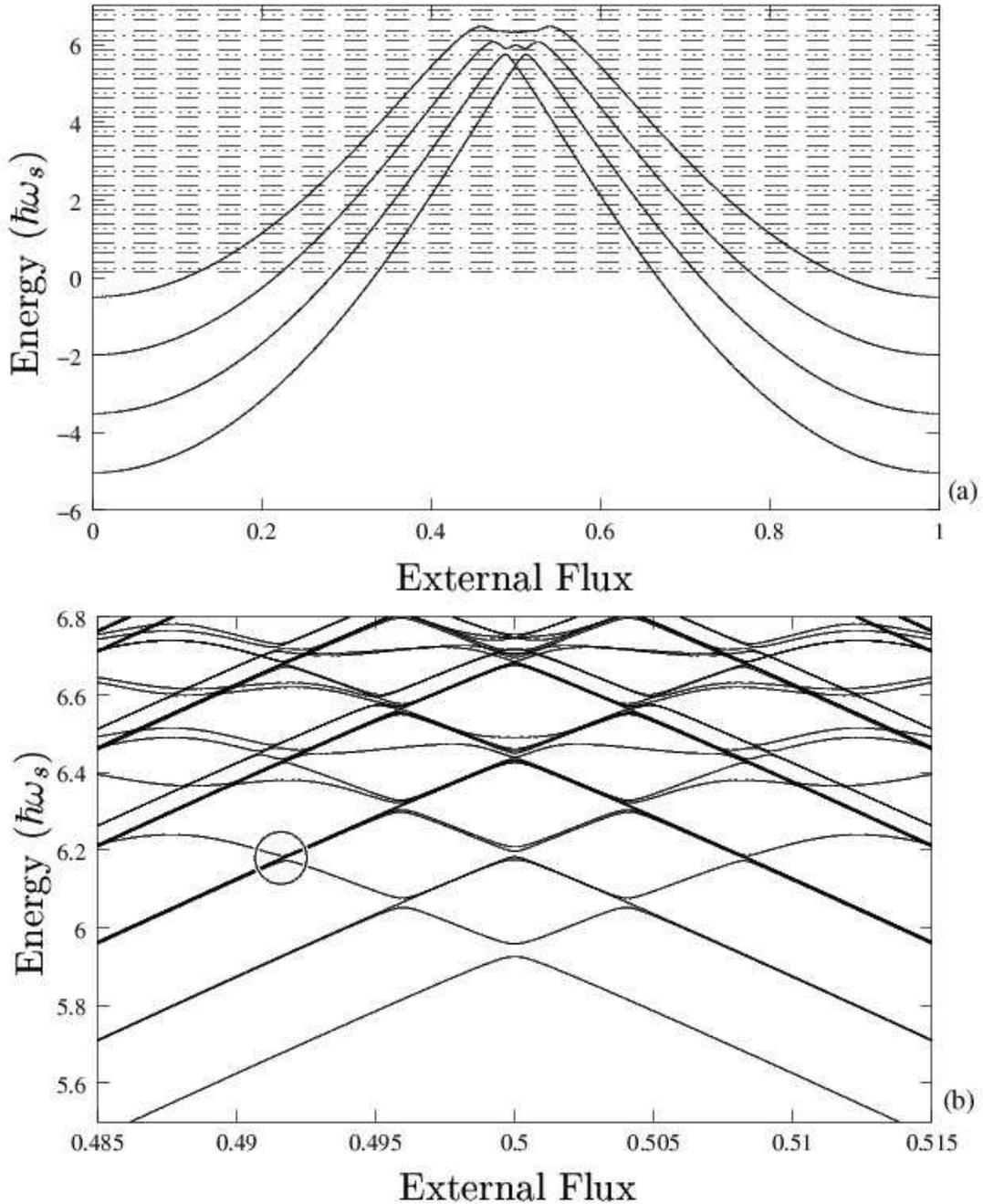}}}
\end{center}
\caption{\textbf{(a)} the eigenenergy spectrum of an uncoupled mesoscopic
SQUID ring at $T=0\mathrm{K}$ as a function of external bias flux (solid
lines) together with the harmonic oscillator spectra for the input mode
(dotted) and the two output modes (dashed) \textbf{(b)} the eigenenergy
spectrum of the coupled system (figure~\protect\ref{QDCFactor2_Schematic})
for the four modes coupled together inductively with the circuit parameters
given in the text.}
\label{QDCFactor2_Spectrum}
\end{figure*}
figure~\ref{QDCFactor2_Spectrum}(a) the first four eigenenergies $\left(
\kappa =0,1,2,3\right) $ as a function of external flux $\Phi _{x}$ for a
typical mesoscopic quantum SQUID ring with parameters~\cite{mje}
$C_{s}=5\times 10^{-15}\mathrm{F}$, $\Lambda _{s}=3\times
10^{-10}\mathrm{H}$ (giving a characteristic SQUID ring
oscillator frequency of $130$GHz, this is much smaller than the
superconducting energy gap in, for example, Niobium, that is often
used to fabricate SQUID circuits) and $\hbar \nu =0.035\Phi
_{0}^{2}/\Lambda _{s}$ equivalent to a Josephson
frequency of $700$GHz, which is not unrealistic when compared
to those used by Friedman et al~\cite{friedman} ($2$THz) and
Orlando et al~\cite{orlando} ($2.12$THz). These frequencies
should be sufficient to avoid thermal excitations for any experiments
performed at $40$mK, a temperature easily obtainable in modern
dilution refrigerators. If we assume that the Josephson weak link in
the SQUID ring is of the tunnel junction type with, typically, an
oxide tunnel barrier of dielectric constant 10, then the junction
dimensions will be close to $0.2\mathrm{\mu m}$ square, which can be
fabricated using current lithographic and thin film
techniques. Furthermore, with $\hbar \nu =0.035\Phi _{0}^{2}/\Lambda
$, and these dimensions, the corresponding Josephson critical current
density is $\approx 4\mathrm{kA/cm}^{2}$ which appears suitable for
SQUID rings operating in the quantum regime. In addition to the ring
eigenenergies, we also show in figure~\ref{QDCFactor2_Spectrum}(a) the
harmonic oscillator field mode eigenenergies of $e_{1}\,\left( \hbar
\omega _{e_{1}}=\hbar
\omega _{s}/2\right) $ (dotted) and $e_{2\text{,}3}\,\left( \hbar
\omega _{e_{2\text{,}3}}=\hbar \omega _{s}/4\right) $ (dashed). In
figure~\ref{QDCFactor2_Spectrum}(b) we show the eigenenergies of the
total coupled system. We note that at the level of capacitance chosen
for the SQUID ring the energy splittings where the field mode energies
and the ring energies cross are very small. In
figure~\ref{QDCFactor2_Spectrum}(b) almost all these splittings are
unresolved on this scale but at sufficient resolution all would be
discernible.

Provided that we disallow any direct interaction between the field mode
oscillators (this being our assumption and a situation which is easy to
establish experimentally), the coupling terms are given by 
\begin{equation*}
H_{e_{i}s}=\frac{\mu_{e_{i}s}}{\Lambda_{s}}\left( \Phi_{s}-\Phi_{x}\right)
\Phi_{e_{i}},
\end{equation*}
where the $\mu_{e_{i}s}$ are the fractions of magnetic flux coupled
inductively between any one of the field mode oscillators and the SQUID
ring. The total system Hamiltonian $H_{t}$ for the three em mode system then
reads 
\begin{equation}
H_{t}=H_{s}+\sum_{i=1}^{3}\left( H_{e_{i}}-H_{e_{i}s}\right)  \label{totham}
\end{equation}

We can make a transformation using the unitary operator $T=\exp\left(
-i\Phi_{x}Q_{s}/\hbar\right) $ to translate the SQUID Hamiltonian $H_{s}$
into a more convenient form. The Hamiltonian (\ref{eq:HamS}) can then be
written as~\cite{mje} 
\begin{equation}
{H}_{s}^{\prime}=T^{\dagger}H_{s}T=\frac{Q_{s}^{2}}{2C_{s}}+\frac{\Phi_{s}
^{2}}{2\Lambda_{s}}-\hbar\nu\cos\left( 2\pi\frac{\Phi_{s}+\Phi_{x}}{\Phi_{0} 
}\right)  \label{eq:HamST}
\end{equation}
where the $\Phi_{0}$-periodic dependence of the (transformed) ring
Hamiltonian on $\Phi_{x}$ is explicit. This unitary transformation
also modifies the ring-field mode interactions to the form
$H_{e_{i}s}^{\prime}=\frac{\mu_{e_{i}s}}{\Lambda_{s}}\Phi_{s}\Phi_{e_{i}}$
so that, following on from (\ref{totham}), the transformed system
Hamiltonian (again $\Phi_{0}$-periodic in $\Phi_{x}$) can the be
written as
\begin{equation}
H_{t}^{\prime}=H_{s}^{\prime}+\sum_{i=1}^{3}\left( H_{e_{i}}-H_{e_{i}
s}^{\prime}\right)  \notag
\end{equation}

Adopting this Hamiltonian we can solve the eigenproblem
$H_{t}^{\prime}\left\vert \xi_{n}\right\rangle =\Xi_{n}\left\vert
\xi_{n}\right\rangle $ and use the eigenvectors $\left\vert
\xi_{n}\right\rangle $ to construct the evolution operator $U(t)$ for the
system~\cite{mje} via, 
\begin{equation}
U\left( t\right) =\sum_{n}\left\vert \xi_{n}\right\rangle \exp\left(
-\frac{i\Xi_{n}t}{\hbar}\right) \left\langle \xi_{n}\right\vert
\label{eq:evolution}
\end{equation}
The time averaged energy expectation values $\left\langle \left\langle
H_{i}\right\rangle \right\rangle $ for the ring and the field modes can then
be calculated from the expression

\begin{equation}
\left\langle \left\langle H_{i}\right\rangle \right\rangle =\underset{\tau\rightarrow\infty}{\lim}\int_{0}^{\tau}\mathrm{Tr}\left[ \rho_{i}\left(
t\right) H_{i}\right] dt \label{eq:timeaven}
\end{equation}
where the $\rho_{i}\left( t\right) $ are the reduced density operators
for the sub-components $i=e_{1},e_{2},e_{3}$ and $s$ for the coupled
system~\cite{cohen} which are obtained from the density operator
$\rho\left( t\right) =U(t)\rho(0)$ (for an initial density operator
$\rho(0)$) by using $\rho_{i}\left( t\right) =\mathrm{Tr}_{j\neq
i}\left\{ \rho\left( t\right)
\right\} $.

\subsection{Four mode (two photon) down conversion and entanglement}

\subsubsection{Without dissipation}

Referring again to the system of figure~\ref{QDCFactor2_Schematic}, and
assuming that it starts in an initial state $\left\vert \psi\left( 0\right)
\right\rangle $, the evolution operator (\ref{eq:evolution}) can then be
used to generate the system state vector $\left\vert \psi\left( t\right)
\right\rangle =U(t)\left\vert \psi\left( 0\right) \right\rangle $ at any
subsequent time $t$ (from which $\rho(t)$ can be found via
$\rho(t)=\left\vert \psi\left( t\right) \right\rangle \left\langle
\psi\left( t\right) \right\vert $). As an illustration we now assume
that the system, SQUID ring + oscillator modes, starts in the state
$\left\vert \psi\left( 0\right) \right\rangle =\left\vert
1\right\rangle _{e_{1}}\otimes\left\vert
\alpha\right\rangle _{s}\otimes\left\vert 0\right\rangle _{e_{2}}
\otimes\left\vert 0\right\rangle _{e_{3}},$ where $\alpha$ is the ground
state of the ring and the $\left\vert n\right\rangle _{e_{i}}$ are the
photon number states for the various field modes of the system.

Starting in this initial state we show in 
\begin{figure}[!t]
\begin{center}
\resizebox*{0.45\textwidth}{!}{\includegraphics{{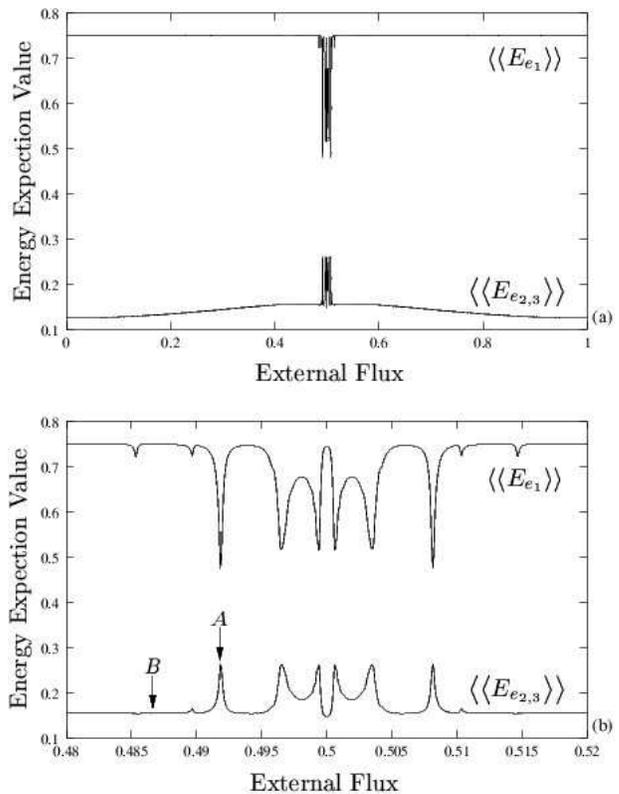}}}
\end{center}
\caption{\textbf{(a)} the time averaged energy expectation values for the
electromagnetic input and output modes of the coupled system of
figure~\protect\ref{QDCFactor2_Spectrum} as a function of external
bias flux
\textbf{(b)} an zoomed in enlargement of the central section of the results
in (a) as a function of bias flux showing a series of energy exchange
regions; in the calculations presented subsequently the two bias points A
and B have been selected, the former inside an exchange region, the latter
outside.}
\label{QDCFactor2_TAE}
\end{figure}
figure~\ref{QDCFactor2_TAE}(a) the time averaged energy expectation
values of the em field modes $\left( e_{1}\text{ - the input and
}e_{2,3}\text{ - the two outputs}\right) $ of the system as a function
of $\Phi_{x}$, where the two outputs $\left( e_{2}\text{ and
}e_{3}\right) $ superimpose exactly, as is to be expected by
symmetry. As is apparent, due to the interaction of the SQUID ring
with $e_{1}$, and one of the other two (output) modes $e_{2}$ and
$e_{3}$, sharp transition regions (or energy exchange
regions~\cite{mje}) develop in the time averaged energies, discussed
by us in a previous publication~\cite{mje}. These regions of energy
exchange between the ring and the field modes occur over very narrow
ranges in $\Phi_{x}$, which is clear from our example system in
figure~\ref{QDCFactor2_TAE}(a) and (b). As we shall demonstrate, the
non-perturbative nature of the ring in these exchange regions is
sufficient to generate photon down conversion and entanglement between
the field modes. In an earlier calculation~\cite{mje} we showed that
ring-field mode interaction strength is a maximum at the centre of the
exchange region and very small at its edge. For the calculations
presented in this section we have chosen to bias the coupled system at
the centre of the exchange region, denoted $A$ in
figure~\ref{QDCFactor2_TAE}(b). The energy exchange region is seen as
the energy splitting (divided crossing) circled in
figure~\ref{QDCFactor2_Spectrum}(b).

\begin{figure}[t]
\begin{center}
\resizebox*{0.45\textwidth}{!}{\includegraphics{{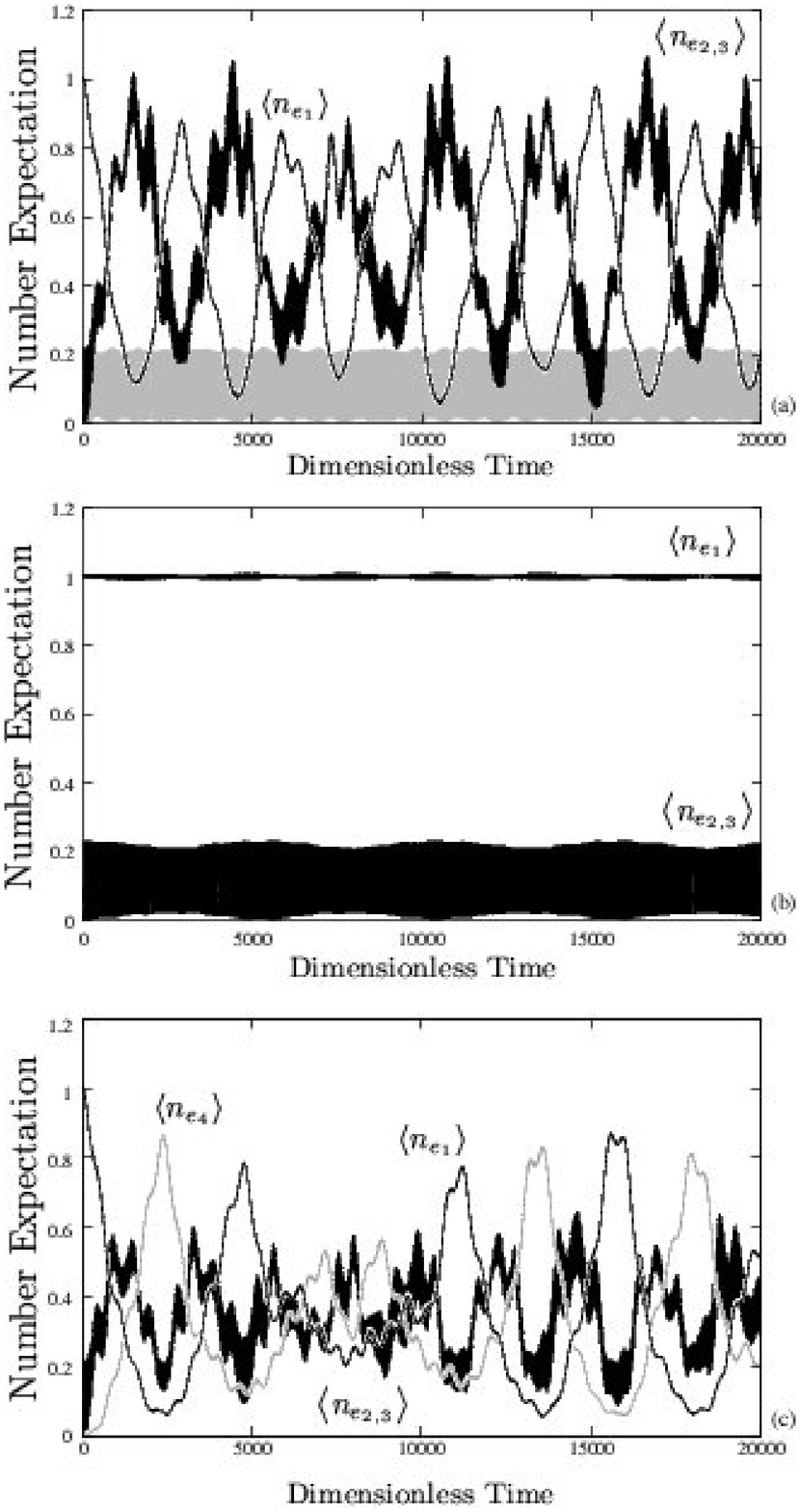}}}
\end{center}
\caption{\textbf{(a)} number expectation values for the electromagnetic
input and output modes of the coupled system of
figure~\protect\ref{QDCFactor2_Spectrum} as a function of
dimensionless time, at external flux point A in
figure~\protect\ref{QDCFactor2_TAE}, starting with a photon occupancy
of 1 in the input mode; also shown in grey is the output mode number
expectation values for the zero photon occupancy in the input mode
\textbf{(b)} the same results as (a) above calculated at the external bias
flux point B in figure~\protect\ref{QDCFactor2_TAE}. \textbf{(c)} a repeat
of the calculation of (a) with the inclusion of a third output mode $e_{4}$,
with its number occupancy shown in grey, at the same frequency as our input
mode.}
\label{QDCFactor2_Number}
\end{figure}

In figures~\ref{QDCFactor2_Number}(a) and (b) 
we show how the number expectation values $\left\langle
n_{e_{1}}\right\rangle $, $\left\langle n_{e_{2}}\right\rangle $ and
$\left\langle n_{e_{3}}\right\rangle $ of the three em field modes
vary with the dimensionless time $\left( \tau =\omega _{s}t\right) $
for the bias points $A$ and $B$ in figure~\ref{QDCFactor2_TAE}(b),
where $\Phi _{x}=0.49183\Phi _{0}$ (at A in
figure~\ref{QDCFactor2_Number}(a)) and $\Phi _{x}=0.487\Phi _{0}$ (at
$B$ in figure~\ref{QDCFactor2_Number}(b)). Here, we note first that
for bias point $A$ there is down conversion of energy from field mode
$e_{1}$ to one or other of the two half frequency $\left( \omega
_{e_{1}}/2\right) $ modes $e_{2,3}$ and second that $\left\langle
n_{e_{2}}\right\rangle $ and $\left\langle n_{e_{3}}\right\rangle $
versus $\omega _{s}t$ are identical due to the symmetric nature of
their coupling to the SQUID ring. In order to verify that the photon
occupation of the $e_{2,3} $ modes is due to down conversion from mode
$e_{1}$, as stated above, we repeated our calculation at bias point
$A$ with the system in the initial state $\left\vert \psi \left(
0\right) \right\rangle =\left\vert 0\right\rangle _{e_{1}}\otimes
\left\vert \alpha \right\rangle _{s}\otimes
\left\vert 0\right\rangle _{e_{2}}\otimes \left\vert 0\right\rangle _{e_{3}}$. 
The time averaged energies $\left\langle n_{e_{2}}\right\rangle $ and
$\left\langle n_{e_{3}}\right\rangle $ for this situation are shown in
grey in figure~\ref{QDCFactor2_Number}(a). As expected, with $e_{1}$
starting in state $\left\vert 0\right\rangle _{e_{1}}$there was no
down conversion.  However, we note that there is some occupancy in the
output modes due to the ring-mode coupling energy available in the
system. It is, of course, possible that higher harmonics of the
output field oscillators could be excited, which may not be
desirable. In order to investigate such excitations we have included
an additional output mode, at the same frequency as the input mode,
and recomputed the results of figure~\ref{QDCFactor2_Number}(a).  This
is shown in figure~\ref{QDCFactor2_Number}(c) Although it is apparent
that the level of down conversion is reduced it is, however, still
significant. It would appear, therefore, that the presence of unwanted
higher harmonics in the output modes is a factor which might need to
be reduced. Nevertheless, it does not negate the computations where
such processes are not considered.

Concomitant with the two photon down conversion $e_{1}\longrightarrow
e_{2,3} $ there exists the possibility that the photons in the lower
frequency modes $e_{2}$ and $e_{3}$ are strongly correlated or even
entangled. In order to characterise the correlations within this
system we will employ two entropic quantities. The first of these
yields information about the correlations that exist between any two
components of a
system~\cite{linbland,leib1970,Wehrl1978,BarnettP1991,mje} and is
defined by
\begin{equation}
I\left( A,B\right) =S_{A}+S_{B}-S_{AB}  \label{2,3entang}
\end{equation}
where $S_{m}$ is the von Neumann entropy for each subsystem $\left( m\right) 
$~\cite{nielsen}. We note that for a bipartite closed system this measure is
often employed to characterise the entanglement present. However,
complications arise when this quantity is used in multipartite or open
systems (for more information see~[\onlinecite{scott}]). Nevertheless, it is
still useful for gaining an appreciation of the correlations that are
present between the components of the system in which we are interested. The
second entropic quantity we shall use is the entanglement measure introduced
by Adami and Cerf~\cite{cerf}, which for a subsystem $A$ of the total system $T$
is given by 
\begin{equation}
E(A)=S_{T}-S_{A}  \label{eq:cerf}
\end{equation}
This measure has the important property that if it is negative definite then 
$A$ is quantum correlated (i.e. entangled) with at least some of the rest of
the system.

For this particular example we are interested in the correlations that
exist between the two output mode, $e_{2}$ and $e_{3}$. The computed
entropy, $I\left(
e_{2},e_{3}\right)=S_{e_{2}}+S_{e_{3}}-S_{e_{2},e_{3}}$, is shown in
\begin{figure}[t]
\begin{center}
\resizebox*{0.45\textwidth}{!}{\includegraphics{{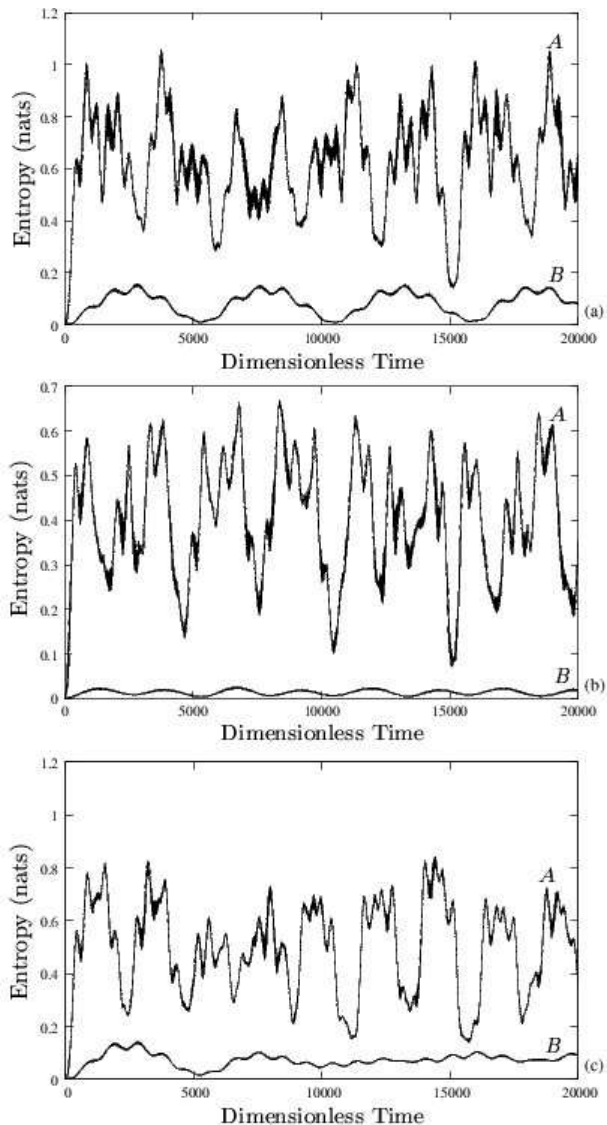}}}
\end{center}
\caption{\textbf{(a)} the entanglement entropy as function of dimensionless
time between the two electromagnetic output modes of the system of
figure~\protect\ref{QDCFactor2_Spectrum} calculated at the bias flux
points $A$ and $B$ in
figure~\protect\ref{QDCFactor2_TAE}(b). \textbf{(b)} the entanglement
entropy between the input mode and one of the output modes as a
function of dimensionless time, again calculated at the two bias
points $A$ and $B$ in
figure~\protect\ref{QDCFactor2_TAE}. \textbf{(c)} a repeat of the
calculation of (a) modified by the inclusion of a third output mode at
the same frequency as our input mode.}
\label{QDCFactor2_Entanglement}
\end{figure}
figure~\ref{QDCFactor2_Entanglement}(a) as a function of dimensionless
time $\omega _{s}t$ for the bias points $A$ and $B$ in
figure~\ref{QDCFactor2_TAE}(b). In addition, in
figure~\ref{QDCFactor2_Entanglement}(b) we show the individual, and
identical, entropies of the two (receiver) modes $e_{2}$ and $e_{3}$
with the initial em field mode $e_{1}$, i.e. $I\left(
e_{1},e_{i}\right) =S_{e_{1}}+S_{e_{i}}-S_{e_{1},e_{i}}$. $i=2$ or $3$
for the same two bias points $A$ and
$B$. Figure~\ref{QDCFactor2_Entanglement}(c) shows a repeat of the
calculation of figure~\ref{QDCFactor2_Entanglement}(a) having included
an additional output mode at the same frequency as the input mode, as
in figure~\ref{QDCFactor2_Number}(c). Again, the inclusion of this
extra output mode has modified the results. However, the two low
frequency output modes ($e_{2}$ and $ e_{3}$) are still significantly
entangled with each other at the appropriate point in external flux,
again indicating that the excitation of additional output modes
reduces, but does not suppress, the entanglement in between the two
low frequency output modes ($e_{2}$ and $e_{3}$). We note that the use
of an actual $LC$~\cite{jmodopt,Wallraff2004} circuit would altogether
remove the need to consider higher order harmonics.

\begin{figure}[t]
\begin{center}
\resizebox*{0.45\textwidth}{!}{\includegraphics{{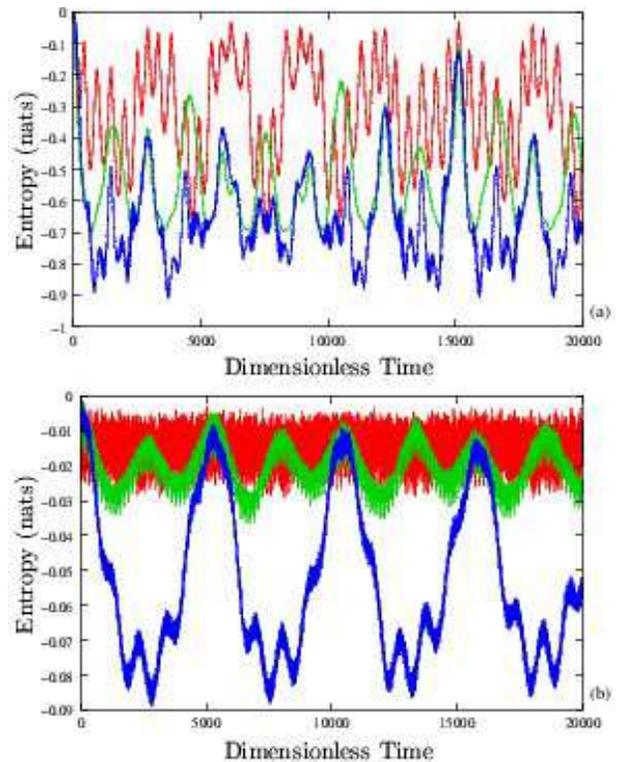}}}
\end{center}
\caption{\textbf{(a)} the Adami-Cerf entanglement entropies for the coupled
system of figure~\protect\ref{QDCFactor2_Spectrum} as a function of
dimensionless time calculated at point $A$ in
figure~\protect\ref{QDCFactor2_TAE}, where the SQUID ring results are
in red, the input electromagnetic mode results are in green and those
for the two output modes are in blue \textbf{(b)} the same calculation
as above in (a) but now at the external bias flux point $B$ in
figure~\protect\ref{QDCFactor2_TAE}.}
\label{QDCFactor2_AdamiCerf}
\end{figure}

In the photon number expectation values of
figure~\ref{QDCFactor2_Number} it is apparent that the decrease in
number expectation value of field mode $e_{1}$ is directly linked to
its increase in modes $e_{2}$ and $e_{3}$ and that the degree of
transfer is highly dependent on the value of $\Phi _{x}$ in the
exchange region. Correspondingly, from
figure~\ref{QDCFactor2_Entanglement}(a) it is clear that, with the
SQUID ring biased into an exchange region, the correlations between
the $e_{2}$ and $e_{3}$ modes are greatly affected by the choice of
$\Phi _{x}$. Thus, at $B$ the correlations are relatively close to
zero at all times while at $A$ it has a maximum of around 1.1. This
$\Phi _{x}$-dependent variation in the strength of correlation is also
seen in figure~\ref{QDCFactor2_Entanglement}(b) between the $e_{1}$
and $e_{2}$ (or equivalently $e_{3}$) receiver mode.  Again these
correlations between each pair of coupled modes is identical, whether
we choose bias point $A$ or $B$. In order to characterise the quantum
correlations with this system we show in
figure~\ref{QDCFactor2_AdamiCerf} the Adami-Cerf entanglement
entropies for each mode of the system where, in
figure~\ref{QDCFactor2_AdamiCerf}(a), for bias point $A$ in
figure~\ref{QDCFactor2_TAE}(b), all four modes are entangled with the
rest of the system throughout the evolution while in
figure~\ref{QDCFactor2_AdamiCerf}(b), while at bias point $B$, it is
clear that the level of entanglement is much less. It is apparent in
these figures that both the von Neumann and Adami-Cerf approaches
indicate that the level of entanglement of the modes in the total
system can be controlled by adjusting the external bias flux on the
SQUID ring. The control over the entanglement which is manifest in
figure~\ref{QDCFactor2_AdamiCerf} is to be compared to the case in
quantum optics. Here, the optical media used are typically weakly
(polynomially) non-linear as opposed to the extremely strong
non-linear (non-perturbative) properties of the SQUID ring in an
exchange region. Furthermore, these media generally couple weakly to
em fields which does not have to be the case for SQUID rings. Most
importantly, the strength of the ring-field interaction (with the
resulting two photon down conversion and entanglement) in an exchange
region can be varied by adjusting the bias flux on the ring. To the
best of our knowledge, such control is very difficult to achieve with
optical media.

\subsubsection{With dissipation}

As has been discussed at great length in the literature~\cite{weiss}
coupling of a quantum object to classical dissipative environments leads to
decoherence over some characteristic time period. Clearly this is of great
importance for any realistic discussion of entanglement in multipartite
quantum systems. In order to model the effects of dissipation we have
adopted the standard approach, very often used in the field of quantum
optics. In this model the components of the system are coupled to decohering
thermal baths. The master equation for the evolution of the density of the
multi-component systems then has the form~\cite{weiss} 
\begin{widetext}
\begin{equation}
\frac{\partial\rho}{\partial t}=-\frac{i}{\hbar}\left[  H_{t}^{\prime}
,\rho\right]  +\sum_{sc}\frac{\gamma_{sc}}{2\hbar}\left(  M_{sc}+1\right)
\left(  2a_{sc}\rho a_{sc}^{\dag}-a_{sc}^{\dag}a_{sc}\rho-\rho a_{sc}^{\dag
}a_{sc}\right)  +\frac{\gamma_{sc}}{2\hbar}M_{sc}\left(  2a_{sc}^{\dag}\rho
a_{sc}-a_{sc}a_{sc}^{\dag}\rho-\rho a_{sc}a_{sc}^{\dag}\right)
\label{masterequ}
\end{equation}
\end{widetext}
where $M_{sc}$ is related to the temperature $T=4.2\mathrm{K}$, and
frequency $\omega _{bsc}$, of each decohering bath for each
subcomponent $\left( sc\right) $ of the system via $M_{sc}=\left( \exp
\left( \hbar \omega _{bsc}/k_{B}T\right) -1\right) ^{-1}$. 
In equation~(\ref{masterequ}) $\gamma _{sc}$ is the coupling (i.e. the
damping rate) between each of the system subcomponents and its
respective bath. The choice to couple each subcomponent to a separate
decohering bath is in keeping with the standard decoherence model used
for open quantum systems~\cite{weiss}.  There is, of course, the
possibility that in certain situations all the elements of this system
could couple to a mutual decohering bath~\cite{yu}. This is an
interesting scenario that would require a detailed analysis beyond the
scope of this paper.

\begin{figure}[!b]
\begin{center}
\resizebox*{0.45\textwidth}{!}{\includegraphics{{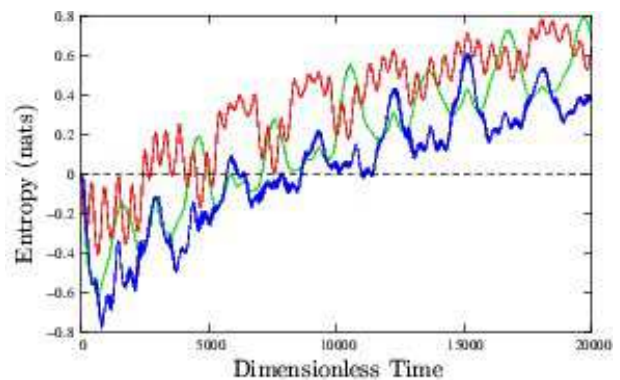}}}
\end{center}
\caption{Repeat of the calculation in figure~\protect\ref{QDCFactor2_AdamiCerf}(a) 
but now with dissipation introduced for a
$\protect\gamma=10^{-5}\protect\omega_{s}$ with the colour coding
green (input mode), blue (output mode) and red (SQUID ring).}
\label{QDCFactor2_ACDissipation1}
\end{figure}

\begin{figure}[!b]
\begin{center}
\resizebox*{0.45\textwidth}{!}{\includegraphics{{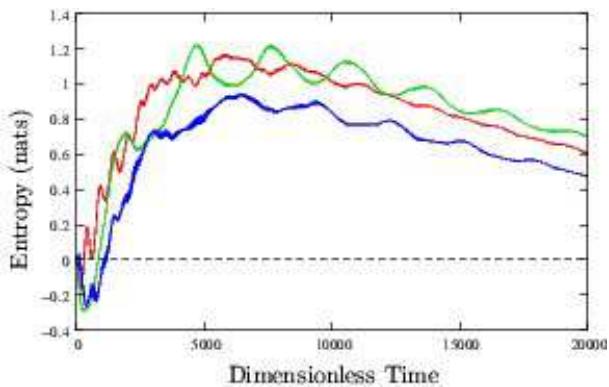}}}
\end{center}
\caption{Repeat of the calculation in figure~\protect\ref{QDCFactor2_AdamiCerf}(a) 
but now with dissipation introduced for a
$\protect\gamma=10^{-4}\protect\omega_{s} $ with the colour coding
green (input mode), blue (output mode) and red (SQUID ring).}
\label{QDCFactor2_ACDissipation2}
\end{figure}

Following on from the entanglement entropy versus dimensionless time
plots for the four mode system in the absence of dissipation
(figure~\ref{QDCFactor2_AdamiCerf}), we show in
figures~\ref{QDCFactor2_ACDissipation1}
and~\ref{QDCFactor2_ACDissipation2}
the subcomponent (Adami-Cerf) entropies for two levels of dissipation,
namely $\gamma=10^{-5}\omega_{s}$ and $\gamma=10^{-4}\omega_{s}$. As
can be seen, in the less dissipative case shown in
figure~\ref{QDCFactor2_ACDissipation1} (which would appear to be
attainable experimentally~\cite{lupascu,martinis}) entanglement is
maintained over a relatively long time period. However, as one would
expect, over long enough times any original entanglement
disappears. Nevertheless, for the case of
figure~\ref{QDCFactor2_ACDissipation1} it appears that a useful level
of entanglement is maintained amongst the output modes over a range of
over 5000 in normalised time or $3\times10^{-8}\mathrm{s}$ in real
time. This is considerably longer than the time constant corresponding
to the $\Lambda_{s}C_{s}$ oscillator frequency of the SQUID ring,
i.e. $\left( 1/2\pi\sqrt{\Lambda_{s}C_{s}}\right) ^{-1}
=7\times10^{-12}\mathrm{s}$ for
$\Lambda_{s}=3\times10^{-10}\mathrm{H}$ and
$C_{s}=5\times10^{-15}\mathrm{F}$. Thus, provided the level of
dissipation can be kept at or below this level it should prove
possible to utilise this two photon entanglement in practical
situations.

\section{Six mode (four photon) down conversion}

\begin{figure}[!t]
\begin{center}
\resizebox*{0.45\textwidth}{!}{\includegraphics{{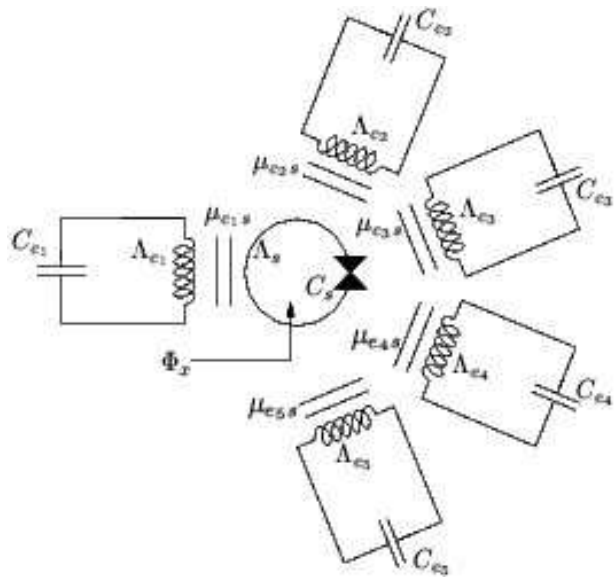}}}
\end{center}
\caption{Schematic for a six mode system with a mesoscopic quantum regime
SQUID ring inductively coupled to one input and four output electromagnetic
modes.}
\label{QDCFactor4_Schematic}
\end{figure}

\begin{figure}[!t]
\begin{center}
\resizebox*{0.45\textwidth}{!}{\includegraphics{{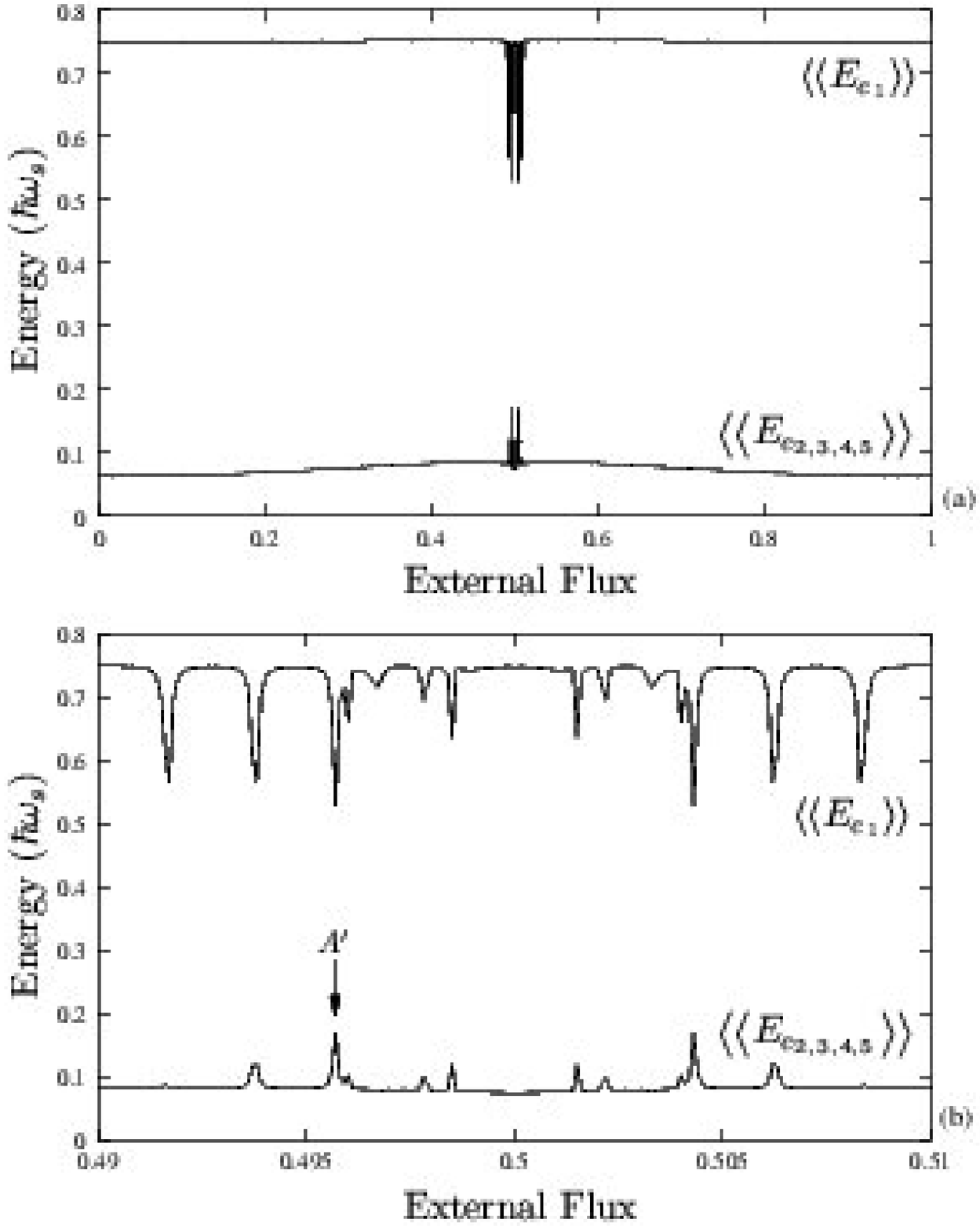}}}
\end{center}
\caption{\textbf{(a)} the time averaged energies at $T=0\mathrm{K}$ for the
electromagnetic input and output modes of the coupled system of
figure~\protect\ref{QDCFactor4_Schematic} as a function of external
bias flux with the circuit parameters given in the text \textbf{(b)} a
zoomed in enlargement of the central section of the results in (a) as
a function of bias flux showing a series of energy exchange regions;
in the calculations presented subsequently the bias point $A^{\prime}$
has been selected.}
\label{QDCFactor4_TAE}
\end{figure}

\begin{figure}[!t]
\begin{center}
\resizebox*{0.45\textwidth}{!}{\includegraphics{{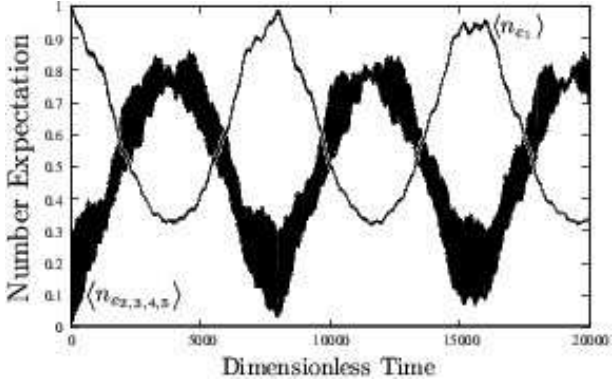}}}
\end{center}
\caption{Number expectation values for the electromagnetic input and output
modes of the coupled system of figure~\protect\ref{QDCFactor4_Schematic} as
a function of dimensionless time, at external flux point $A^{\prime}$ in
figure~\protect\ref{QDCFactor4_TAE}, starting with a photon occupancy of 1
in the input mode.}
\label{QDCFactor4_Number}
\end{figure}

The scheme for a six mode system, consisting of one input oscillator mode
(frequency $\omega_{e_{1}}/2\pi$) and four identical output modes, all
oscillating at one quarter of the input frequency, is shown diagrammatically
in figure~\ref{QDCFactor4_Schematic}. 
Using the same procedure as we adopted in the case of the two photon down
conversion described above, we have computed the number expectation values
against normalised time for the input, the SQUID ring and the four output
modes. These have been computed at the bias flux point $A^{\prime}$ in the
centre of the exchange region (divided crossing) of figure~\ref{QDCFactor4_TAE}. 
These expectation values are plotted in figure~\ref{QDCFactor4_Number} 
for the input mode and the four output modes, each of which generates
identical results. As can be seen, in the absence of dissipation the sum of
the number expectation values for these output modes is reasonably close to
4, indicating that here the down conversion process via the SQUID ring is
quite efficient. We note that the level of efficiency of this down
conversion process decreases as the splitting frequency at the divided
crossing increases. We suspect that this could be corrected if a
sufficiently thorough review of the external flux dependence were to be
undertaken. However, this is impractical to perform given the limitations on
our current computational power.

The computational power required to generate accurate solutions for this
four photon down conversion, and gain a detailed understanding of the
correlations that exist between the components of the system, is extremely
demanding. As with the two photon down conversion case, but even more so,
the computational power at our disposal did not allow us to study the
correlations between the four (identical) output modes of the system. What
we have been able to do is to take any of the output photons, at a quarter
of the frequency of the input photon, and show that each is strongly
entangled with the rest of the system. By appealing to the earlier results
presented in this paper, it is reasonable to assume that the four output
modes are entangled with each other in addition to any entanglement with the
input mode and the SQUID ring. The problem of describing, and quantifying,
entanglement for systems of more than two quantum particles, is still a
matter for serious debate~\cite{kendon,rungta}. However, in a qualitative
way, 
\begin{figure}[!t]
\begin{center}
\resizebox*{0.45\textwidth}{!}{\includegraphics{{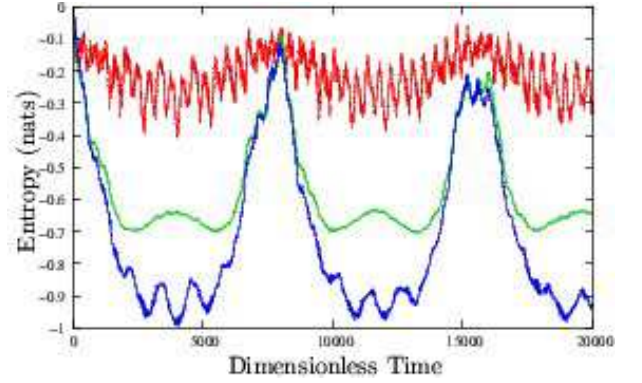}}}
\end{center}
\caption{The Adami-Cerf entanglement entropies of the for the coupled system
of figure as a function of dimensionless time calculated at point
$A^{\prime} $ in figure~\protect\ref{QDCFactor4_TAE}, where the SQUID
ring results are in red, the input electromagnetic mode results are in
green and those for the two output modes are in blue.}
\label{QDCFactor4_AdamiCerf}
\end{figure}
the computed results displayed in figure~\ref{QDCFactor4_AdamiCerf} suggest
a practical route to the entanglement of many identical photon states
through the intermediary of a highly non-perturbative SQUID ring medium.
Again, with regard to the computational limits we faced, we were not able to
introduce dissipation into the many moded system through a master equation
of the type~(\ref{masterequ}).

\section{Higher order down conversion processes}

\begin{figure}[!t]
\begin{center}
\resizebox*{0.45\textwidth}{!}{\includegraphics{{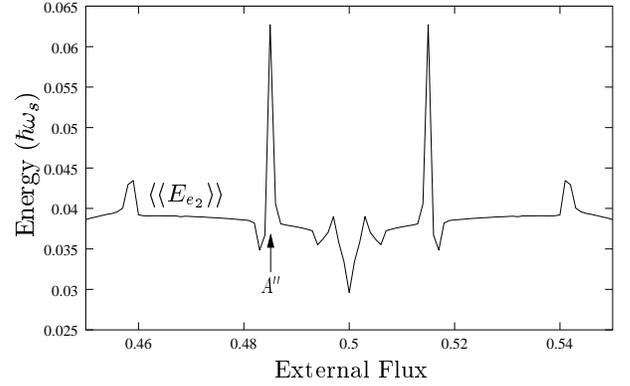}}}
\end{center}
\caption{The $T=0\mathrm{K}$ time averaged energy for the output
electromagnetic mode of a factor 20 SQUID ring mediated down
conversion process as a function of external bias flux where the bias
point $A^{\prime\prime}$ in an energy exchange region is used in
subsequent calculations with the circuit parameters given in the
text. }
\label{QDCFactor20_TAE}
\end{figure}

\begin{figure}[!t]
\begin{center}
\resizebox*{0.45\textwidth}{!}{\includegraphics{{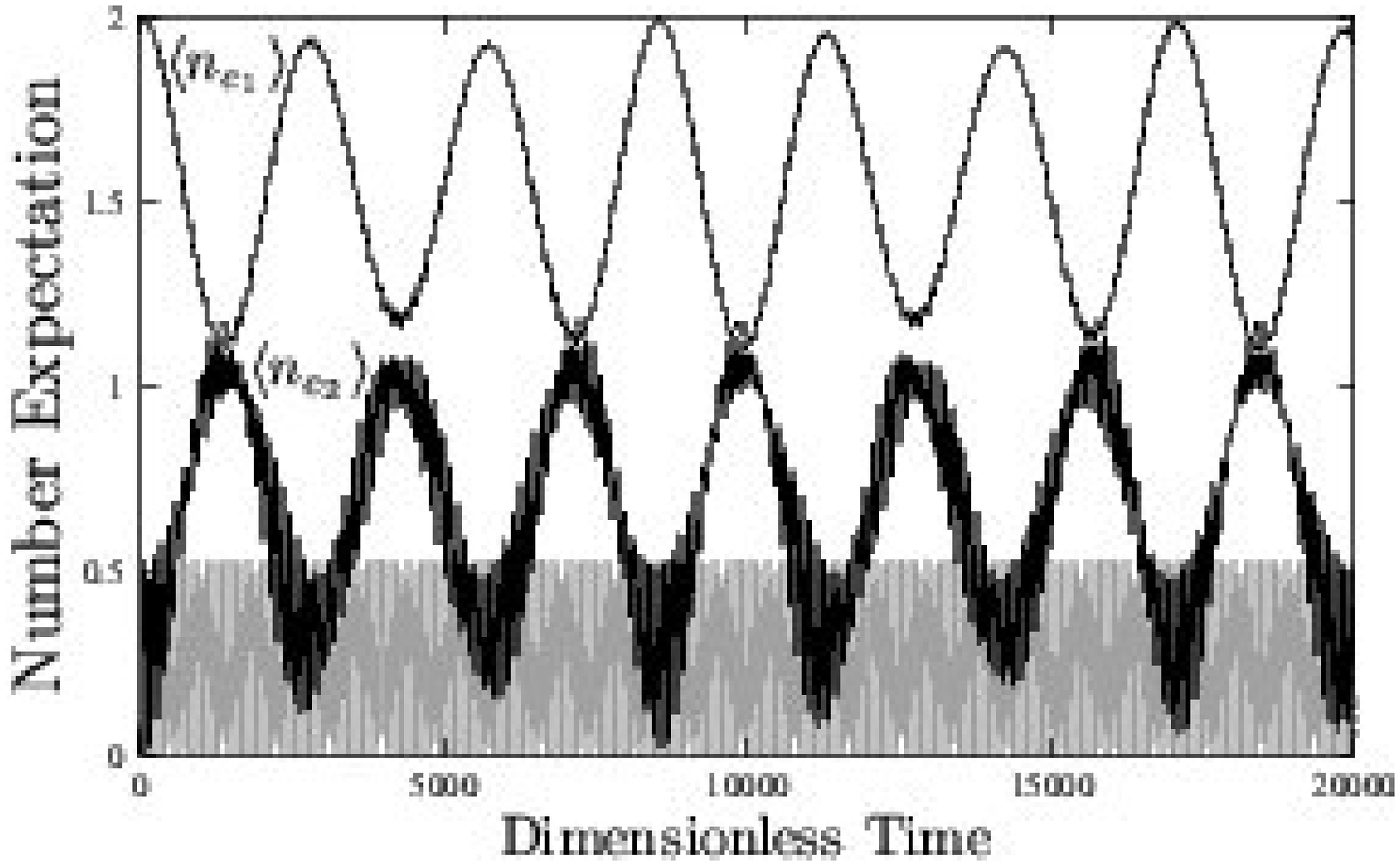}}}
\end{center}
\caption{Number expectation values for the electromagnetic input and output
modes of the coupled system of figure~\protect\ref{QDCFactor20_TAE} as a
function of dimensionless time, at external flux point $A^{\prime\prime}$ in
this figure, starting with a photon occupancy of 2 in the input mode; also
shown in grey is the output mode number expectation values for the zero
photon occupancy in the input mode.}
\label{QDCFactor20_Number}
\end{figure}

Due to the limitations imposed by our computing power we have not been
able to extend our investigations of entanglement beyond the six mode
(four photon down conversion) range. Nevertheless, following the two
and four photon processes we now consider a factor twenty down
conversion process between an input and an output em mode. For this
calculation we have used the same SQUID ring parameters as
before. However, with this calculation describing a factor twenty
frequency conversion, the exchange regions have been redistributed as
shown in figure~\ref{QDCFactor20_TAE}. For the calculation presented
in this section we have chosen to flux bias the ring at the centre of
a particular exchange region in figure~\ref{QDCFactor20_TAE}, denoted
by $A^{\prime\prime}$.
At this value of external flux we have been able to demonstrate
quantum down conversion by a factor of twenty between the input and an
output em oscillator mode. In order to reduce the computational
complexity of this calculation, and in contrast to the previous
examples, we have taken $\omega_{e_{1}} =\omega_{s}$ and assumed an
initial (start) number occupancy in this input mode of $\left\vert
2\right\rangle _{e_{1}}$. In the computation we have inductively
coupled the SQUID ring to an output oscillator mode at one twentieth
of the input photon frequency. Setting the input and output coupling
strengths at $\mu=0.01$, we should see, for a perfectly efficient down
conversion process, that the number expectation value of the output
modes reaches a peak of approximately 20 as a function of normalised
time. The actual level of occupancy in the output mode is much lower
than this potential maximum. However, as discussed earlier, the
efficiency of this down conversion process is greatly affected by the
value of the external bias flux chosen. The results presented in
figure~\ref{QDCFactor20_Number}
are, for the moment, the best that could be obtained within the
practical time and computational constraints imposed on us. In
figure~\ref{QDCFactor20_Number} the number expectation value for the
output oscillator mode $\left( e_{2}\right) $ is plotted as a function
of normalised time.  This confirms that, at least from a theoretical
viewpoint, high order down conversion (and, by implication, up
conversion) processes can occur between input and output em modes
through the intermediary of a quantum regime SQUID ring. It should be
noted that while the occupancy of the lower frequency output mode here
is not very efficient, it is still a non-trivial deviation from
zero. It is therefore still valid to look upon this result as evidence
that SQUID rings in the quantum regime can mediate high order down
conversion, even if, for the moment, at low rates of efficiency. By
extension, if the computational power were available to us, it seems
reasonable to assume that (i) we could obtain the same result if we
were to couple twenty output modes to the SQUID ring and (ii) that
there is every possibility that these output modes would be able to
entangle with each other. Furthermore, given the extremely
non-perturbative nature of a quantum mechanical SQUID ring, we might
expect that, with the appropriate experimental arrangements in place,
down conversion frequency ratios much higher than 20 might be
observed.

\section{Conclusions}

The flexibility afforded by this $\Phi _{x}$-dependent control
(tunability) of the maximum coupling strength, and the temporal form
of the entanglements within ring-field mode(s) system, raises
interesting possibilities. This control might be of great advantage in
the preparation of qubits in quantum computing and quantum information
processing~\cite{lo,bouwmeester}. Our calculations point to a method
for generating a wide range of levels of entanglement between photon
states (or, in general, any quantum circuits coupled to the SQUID
ring) with only very small adjustments in bias flux required. This we
consider to be the important outcome of this work. To give an example,
for bias point $A$ in the data of
figures~\ref{QDCFactor2_Entanglement}(a) and~(b) the time evolution
shows instances, such as at $\omega _{s}t\approx 10,000$, when the
entanglement entropy $I\left( e_{2},e_{3}\right) $ is high and the
entanglement entropy $I\left( e_{1},e_{i}\right) $, $i=2$ or $3$, is
low. Even more significantly, in
figure~\ref{QDCFactor2_ACDissipation1} we see that at dimensionless
times such as $\omega _{s}t\approx 5000$ the output modes are the only
modes which are unequivocally entangled. This would appear to be of
significance in many proposed device
applications~\cite{lo,bouwmeester,orlando,makhlin} where a way is
needed to use a known input to prepare a set of entangled photons and
then to disconnect the input, without significantly disrupting the
entanglement so produced. We note that in many experimental
situations~\cite{linbland,leib1970,Wehrl1978,BarnettP1991} any
readjustment of output entanglement requires the development of a new
system setup. Many of the ideas explored in this work were inspired by
concepts in quantum optics.  In fact, there are strong similarities
between a quantum regime SQUID ring and a device frequently employed
in quantum optics - the non-linear optical coupler - this being formed
from two or more linearly coupled waveguides where at least one is
composed of an optically non-linear
medium~\cite{mogilevtsev,rahacek,fiurasek,armstrong,silberhorn}. However,
there are two important differences between optical couplers and the
SQUID ring: (i) optical non-linear interactions usually have a
non-resonant character and (ii) optical non-linear couplers are
difficult to make because of the strict phase and frequency matching
conditions that must be obeyed by all involved processes
simultaneously. Hence, it may well be that SQUID ring coupled devices,
such as those presented here, could lead to a feasible way of
realising (or at least simulating) non-linear optical couplers.

The manipulation of the different entanglements in the model systems
considered in this paper, based on just one SQUID ring, appears to
open up the possibility of creating more sophisticated, multi-SQUID
ring based, circuit networks where the constituent elements could be
entangled and disentangled as required simply by changing bias
fluxes. Obviously further work is called for to determine how finely
entanglement properties can be manipulated in more complex systems. As
a pointer to the use of such complex systems, we have already shown in
this work that flux controlled, two photon down conversion
($\omega_{1}\rightarrow\omega_{2,3}$ where $\omega_{1}=2\omega_{2,3}$)
and entanglement can occur in a SQUID ring mediated system as can four
photon down conversion ($e_{1}\rightarrow$ $e_{2,3,4,5}$). Given the
cosine nonlinearity of the ring, we would expect the entanglement of
even more photon states to be possible although to demonstrate this is
beyond the limits of the current computational power available to us.

\section{Acknowledgements}

We would like to express our thanks to the Engineering and Physical Sciences
Research Council for the support of this work through its Quantum Circuits
Network Initiative.

\bibliography{references}

\end{document}